\documentclass{article}
\usepackage[utf8]{inputenc}
\usepackage[margin=1in]{geometry}
\usepackage{algorithm}
\usepackage{graphicx}
\usepackage{algpseudocode}
\usepackage{amsmath,amssymb}
\usepackage{color}
\usepackage{natbib}
\usepackage[normalem]{ulem}
\usepackage{hyperref}
\usepackage{rotating}
\usepackage{appendix}

\newcommand{\actual}{\mathcal{D}}
\newcommand{\mock}{\mathcal{M}}
\newcommand{\allplayers}{\mathcal{P}}
\newcommand{\undrafted}{\varphi}
\DeclareMathOperator*{\argmax}{arg\!\max}
\DeclareMathOperator*{\argmin}{arg\!\min}

\title{Improving the Aggregation and Evaluation of NBA Mock Drafts}

\author{\textbf{Jared D. Fisher}\thanks{Contact the corresponding author at fisher@stat.byu.edu. The views expressed in this manuscript are those of the authors and do not necessarily reflect those of our employers.
We would like to thank Richard Yu for his work on the early phases of this project. 
We also thank Nathan Sandholtz, David Grimsman, Chris Archibald, and Nate Hawkins for providing feedback on early versions of this manuscript. 
This work was partially supported by NSF 1745640.
} \\ 
{\normalsize Brigham Young University} \and \textbf{ Colin Montague} \\
{\normalsize Sacramento Kings} }

\date{  \today}

\begin{document}
\maketitle

\begin{abstract}

 %
 Many enthusiasts and experts publish forecasts of the order players are drafted into professional sports leagues, known as mock drafts.
 Using a novel dataset of mock drafts for the National Basketball Association (NBA), we analyze authors' mock draft accuracy over time and ask how we can reasonably use information from multiple authors.  
 To measure how accurate mock drafts are, we assume that both mock drafts and the actual draft are ranked lists, and we propose that rank-biased distance (RBD) of \cite{RBO} is the appropriate error metric for mock draft accuracy. This is because RBD allows mock drafts to have a different length than the actual draft, accounts for players not appearing in both lists, and weights errors early in the draft more than errors later on. 
 We validate that mock drafts, as expected, improve in accuracy over the course of a season, and that accuracy of the mock drafts produced right before their drafts is fairly stable across seasons. 
 To be able to combine information from multiple mock drafts into a single consensus mock draft, we also propose a ranked-list combination method based on the ideas of ranked-choice voting. 
 We show that our method provides improved forecasts over the standard Borda count combination method used for most similar analyses in sports, and that either combination method provides a more accurate forecast over time than any single author.

\end{abstract}

Keywords: expert elicitation, rank-biased overlap, Borda count, ranked-choice voting, instant-runoff voting, rank aggregation

\newpage

\section{Introduction} 

Professional sports leagues often rely on drafts to assign new players to existing teams. Typically, the worse a team was in the preceding season, the earlier in the draft they get to choose players. Drafts are one of only a handful of tools that a sports organization has for gathering players to their team, along with trading players with other teams or signing free-agent players. All of these management decisions depend on player talent as well as anticipating the actions of competing teams. 

Professional sports drafts are perhaps most impactful in the National Basketball Association (NBA), where teams have rosters of only 15 players and only 5 players can play at any given time. Thus, adding a single talented player will have greater impact on an NBA team than teams in other sports where more players participate in each game. The NBA draft consists of teams taking turns choosing players one at a time from a pool of draft-eligible players. The worse a team's performance in the previous season, the better chances they have of having earlier picks. There are two rounds in the draft and 30 teams, which results in 60 players\footnote{Sometimes teams are penalized for detrimental behavior, and one possible penalty is losing one or more draft picks.}  
being picked, and the rest of the players who declared for the draft are then ``undrafted'' but are also free agents to sign contracts with any team. Teams can also trade or swap (rights to) draft picks, and as draft picks are unfettered access to talented players, draft picks are highly valuable assets. Often teams will trade current players in exchange for future draft picks, so some teams will have more or less than the usual two picks in a draft. 

A common forecast in the sports communities is a mock draft, where the author of the mock drafts predicts the order of players chosen in a draft. The goal is not to predict which players will be the best (this is typically called a ``big board'') but to predict the order players are selected in the draft. 
Mock drafts are never perfect, but astute draft experts can combine information from amateur teams, agents, and other sources providing valuable information often not available to the public and occasionally not to all teams either.  Those who seriously study the draft have been dubbed ``draftniks''.  

Naturally, these mocks drafts are read by fans, critiqued in blogs, and  employed by NBA team personnel to inform decisions. However, in this paper, we show that the methodologies typically used for analyzing NBA mock drafts are suboptimal. Specifically, we show there are alternatives for the traditionally-employed error metrics and aggregation methods. We also show that these alternatives are advantageous theoretically and/or empirically.

\section{Defining the Problem}

\subsection{Key assumption} 

A simplistic but reasonable model of the NBA draft would be to assume each NBA team has ranked all available players, and when it is their turn to choose, the team selects the available player ranked highest on their list. So, ideally, we would like to know each team's preferences about players, which would allow us to forecast the draft perfectly (assuming that each team drafts their favorite player who is available). However, given we only get an average of two glimpses at a team's draft board (one pick in each of the first and second round, both of which may be traded to another team), there is not enough data to estimate players' probabilities of being selected at each draft spot without strong assumptions. 

Instead, like many before us, we assume the draft is an ordered/ranked list. Methodologically, assuming the draft is a ranked list allows us to tap into the error metrics and other methods from the literature on errors on rankings and rank aggregation. As per professional basketball, this effectively ignores team-specific preferences and assumes the actual draft reveals the single NBA-wide ordering of players. We feel this is a reasonable assumption, though without getting to glimpse teams proprietary preferences, we may not know how strong an assumption it is.  One glimpse came in 2017, when Boston traded the first pick to Philadelphia for the third pick and other assets. The implication is that Boston preferred Jayson Tatum to the other players, and had Philadelphia and Boston not traded picks in 2017, Jayson Tatum would likely have been the 1st pick of the draft (instead of the third) despite Markelle Fultz being strongly forecasted to be first.  This would be a forecast error of two spots, though under a preferences-based model, it could be viewed as only one spot, or no error at all, given that the Lakers appeared fairly set to choose Lonzo Ball with the second pick.

\subsection{Notation}

Let $\allplayers$ be the set of all available players. Let the actual draft be represented by the mapping $\actual : \allplayers \rightarrow \{1,2,...,|\actual|,\undrafted\}$. Let $\undrafted$ indicate being undrafted, and borrowing from set notation, let $|\actual|$ be the number of non-$\undrafted$ elements, such that $|\actual|$ is the number of players who are drafted. Usually, $|\actual| = 60$. We will represent each mock draft $i$ as a similar mapping $\mock_i: \allplayers \rightarrow \{	1,2,...,|\mock_i|,  \undrafted\}$.  Thinking of a draft as a mapping works well for our purposes. For example, when considering the 2019 NBA Draft, $\actual(\{\text{Zion Williamson}\}) = 1$ and conversely $\actual^{-1}(1) = \{\text{Zion Williamson}\}$. 

\section{An appropriate accuracy metric}

\subsection{Visualizing mock draft errors and their peculiarities}

\begin{figure}
    \centering
    \includegraphics[width=3.2in]{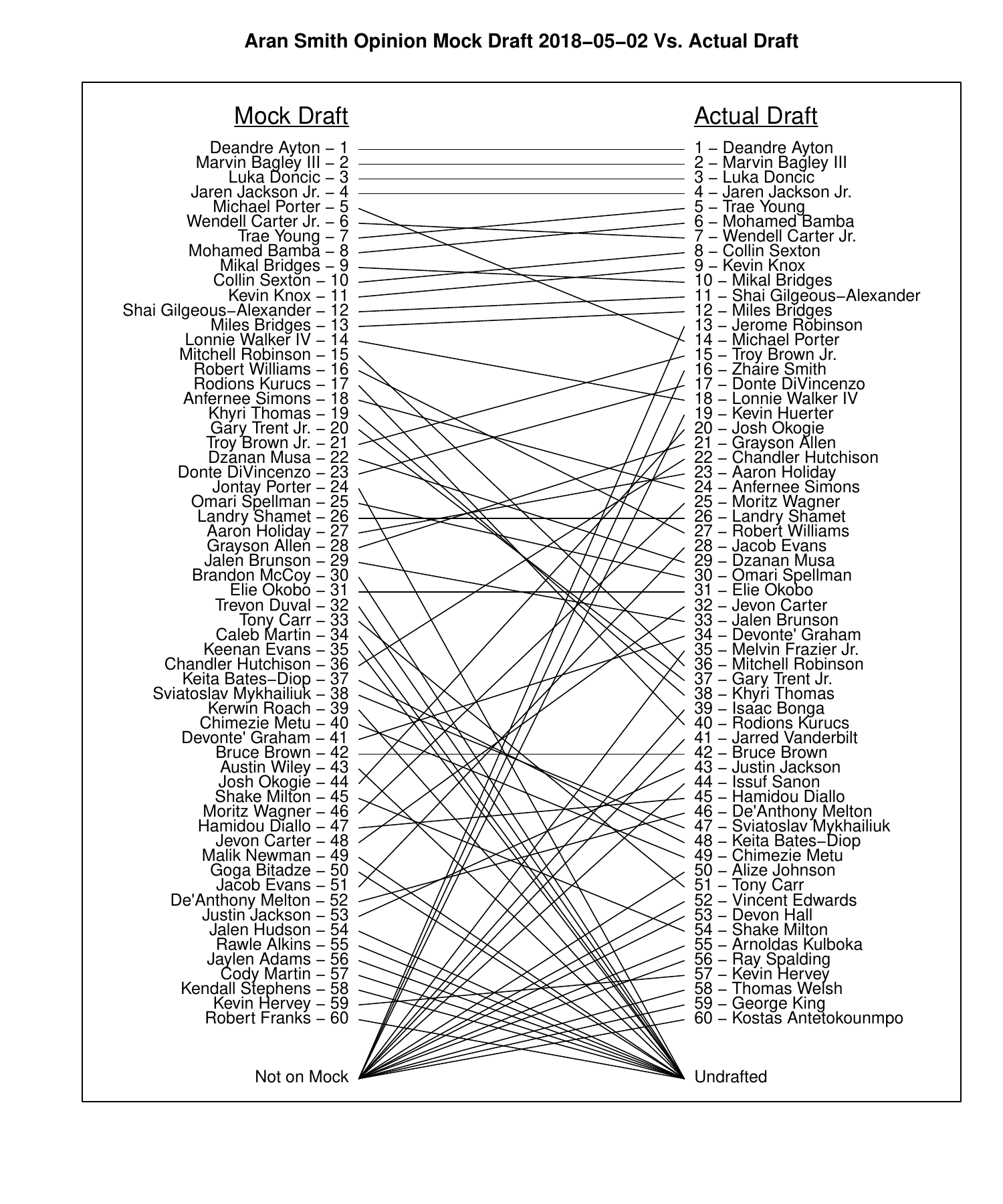}
    \includegraphics[width=3.2in]{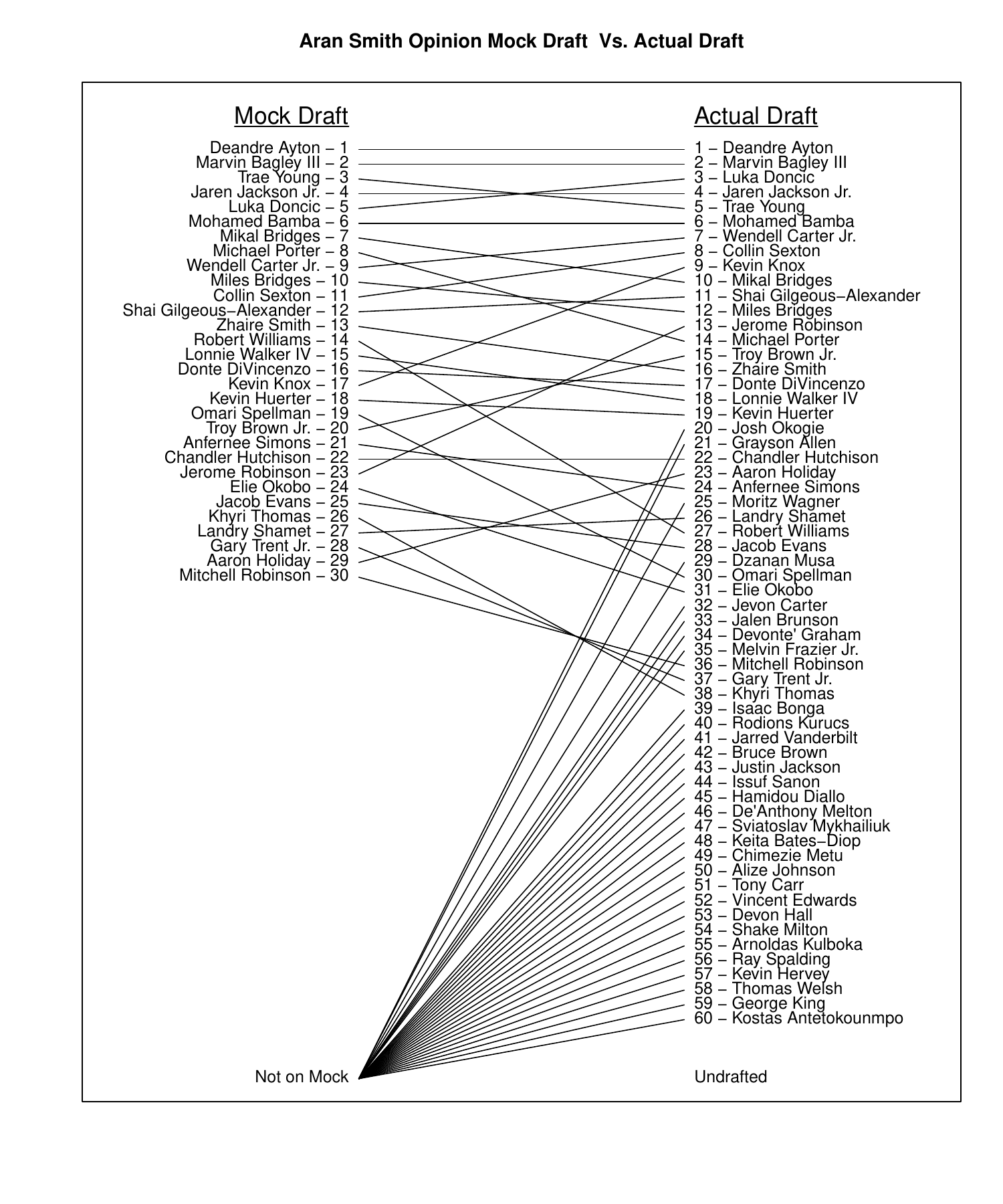}
    \caption{Visualization of mock draft error with slope graphs. Lines connect players' mock draft locations to their spots in the actual draft. Horizontal lines show accurate forecasts, and diagonal lines show forecast errors. The steeper the diagonal line, the more incorrect a author's prediction was. The nodes on the bottom gather the undrafted players included in mock drafts and unmocked players selected in the actual draft.}
    \label{fig:lines}
\end{figure}

The slope graphs in Figure \ref{fig:lines} visualize the accuracy of mock drafts. In each graph, a mock draft is listed on the left side, the actual draft is listed on the right side, and lines connect the draft position of a player from where he was mocked to be drafted and where he was actually drafted. Horizontal lines show accurate predictions, while sloped lines show errors in mocks. The steeper the slope, the further away a player's actual draft position was from his forecasted position. 

Also, there are ``terminal nodes'' on the bottom of each slope graph. As the actual draft will contain a fixed number of players (usually 60), most mock drafts that mock all 60 picks will contain some players that aren't actually drafted. This also means that some players in the actual draft are not included in a given mock draft. 

There are three peculiar features demonstrated in Figure \ref{fig:lines} that we need to keep in mind when evaluating mock draft accuracy.  First, Figure \ref{fig:lines} clearly shows larger errors in the bottom of drafts compared to the top (larger slopes). However, mistakes in forecasting the top of the draft are more important, i.e. little mistakes in the top of the draft are in fact as bad or worse than large mistakes at the end of the draft. 
The distribution of talent among the best players in the world is inherently right-skewed, and as players are selected in order of their perceived talent levels, the talent gap between two players who are drafted one spot apart in the draft will (on average) shrink as the draft progresses. Furthermore, the value of a draft pick is directly tied to the expected talent level of a player that may be selected with that pick. 

Second, while a mock draft may contain anywhere from a handful of players up to about 150, they never include all of the draft eligible players (there are usually 300 or so).
\footnote{Draft eligibility: to play in the NBA, a player must declare for the draft. To be eligible, the player must be at least a certain age and at least a year out of high school. This means that most people on the planet can be eligible, but realistically only the best players from college and non-NBA professional leagues have a chance at being drafted, putting that number well below 500 every year.} 
Neither will the actual draft, even if an extra round or two is added. Furthermore, if produced before the final withdrawal deadline,
\footnote{Withdrawal deadlines: starting with the 2016 NBA Draft, there are two withdrawal deadlines before the NBA draft. The first is specific to the NCAA, and is late May about three weeks before the draft. If players do not opt out before this deadline, they lose their eligibility to play college basketball moving forward as they have opted to become professionals. The second is the general deadline, 10 days before the draft. Players may choose to withdraw before this deadline to stay with their current non-NBA teams, which they may do for a variety of reasons.} 
mock drafts may include players who are not even eligible to be drafted. Thus, both the mock drafts and the actual draft are considered \emph{incomplete}. As incomplete, the two lists will likely not have all the same names, like we see on the left panel of Figure \ref{fig:lines}. This impacts some of the mathematical choices we must make: for example, we have to carefully consider the ``error'' incurred by projecting a player to be drafted 50th but is actually undrafted. This is why neither of the bottom nodes has a numerical value. We do not know where this player would have been drafted had the actual draft been complete, nor do we know who would have been drafted 61st overall. One could consider this missing/censored data. 

Finally, the right panel of Figure \ref{fig:lines} shows a mock draft that only mocked the first 30 picks instead of all 60. This is quite common actually. We don't want to dismiss mock drafts if they do not have 60 players. A common ranking tool is a ``top 100'' big board, and we want to be able to evaluate if these tools are accurate too. Thus, we want to be able to deal with mock drafts of \emph{varying length}.

\subsection{Literature on evaluating the accuracy of mock drafts}

Most public evaluations of mock draft accuracy occurs in the online blogs of sports enthusiasts, and largely for the National Football League. A popular approach is to create a point system, where different amounts of points are given for correct forecasting different parts of the draft.  \cite{fantasypros} use a points-based system with multiple categories. Not only do they give points for minimizing distance from actual draft spot, they give points for correct order within position, if a team drafted that position in the first round, and if the team drafted that player in any spot. Thus, each author gets a certain number of points for each pick in the draft. They grade on the first round of the draft only.  \cite{sportingnews} uses a simpler version of a multi-category point system, giving a full point for a player mocked to go to the correct team with the correct pick, or half points if the player is mocked for the correct pick or the correct team. These are reasonable approaches, however the point values given for each activity are fairly arbitrary, and the points associated with position reflect the heavier importance in the NFL of positional fit within each team. 
 
Many use or adapt standard error metrics. While both \cite{Insider} and \cite{Samford} consider point-based metrics, they also use the mean absolute error between the actual draft position number and the mocked draft position. One hiccup to consider when using such a metric is what to do with undrafted players: if I mock an NBA player to be chosen 55th, and he goes undrafted, what number greater than 60 do I put into my calculation? \cite{Harvard} addresses this by using 61 for the draft position of undrafted players. The methodological challenge is that only one player could be 61st should the draft go that long, and chances are that there are more than one undrafted player in a mock draft of all 60 picks.  

The big issue with standard error metrics based on absolute- or squared-error loss is that they weigh errors early in the draft the same as errors at the end of the draft. However teams (the actual agents/decision makers here) would much rather know exactly how the first round of the draft will play out than the last round. Many of the aforementioned analyses only look at the first round, and this could also be a nod to the need to front-weight mock draft errors. One approach to addressing the weighting problem is to add logarithms into the standard error metrics, as \cite{Robinson} does in his visualizations, which gives mistaking the 1st and 3rd picks the same penalty as the 10th and 30th picks, i.e. $log(1)-log(3) = log(10)-log(30)$.  

\cite{Caudill} address the rank-weighting issue by employing the salary of each draft position. As earlier draft picks are given larger initial salaries than later picks, at the decrease is exponential. So, instead of a players' draft pick number being used in the error metric, the salary of each slot is used, resulting in larger errors for being off one pick early in the draft as opposed to later in the draft. \cite{Caudill} look at the first round of the NFL draft, but we are interested in the entire NBA draft. The salaries for the players chosen in the second round of the NBA draft would not work so well with a salary-based error metric as contracts are actually negotiated and players are sometimes drafted "out of order" because of their willingness to meet certain financial/roster benefits. 

Though good work has been done on this problem in the sports analytics literature, we will look to the literature of other disciplines for solutions better tailored to this exact problem. 
        

As discussed in regards to Figure \ref{fig:lines}, there are three criteria we require for an ideal metric. We clarify these here. 
\begin{enumerate}
    \item Front-weighted: the metric should given more weight to errors early in the draft and down-weight errors later in the draft. Let $\mock_j$, $\mock_k$, $\mock_l$ be mock drafts with the same errors and only two differences: $\mock_k$ has an additional error early in the draft at pick $a$, and $\mock_l$ has an additional error late in the draft at pick $b$, such that $a < b$. Precisely, 
    \begin{align*}
        \mock_k^{-1}(a) \ne \actual^{-1}(a) =  \mock_j^{-1}(a)  =  \mock_l^{-1}(a) \\
        \mock_l^{-1}(b) \ne \actual^{-1}(b) =  \mock_j^{-1}(b)  =  \mock_k^{-1}(b).
    \end{align*}
    Then to be front-weighted, the error metric should hold such that 
    \begin{equation}
        error(\actual,\mock_j)  < error(\actual,\mock_l) < error(\actual,\mock_k)
    \end{equation}
    where the first inequality should be obvious ($\mock_k$ and $\mock_l$ have all the errors of $\mock_j$ and more). The second inequality shows front-weighting: errors earlier in the draft are weighted more in the error metric than errors later in the draft. 
    
    \item Incomplete: the metric should allow for the fact that neither the draft nor the mock drafts will rank all of the players in $\allplayers$: $|\actual| < |\allplayers|$ and $|\mock_i| < |\allplayers|$, $\forall i$. The implication is that the metric can wisely handle undrafted and/or unmocked players, $\undrafted$. 
    \item Unequal lengths: the metric should allow for reasonable measurement of mock drafts where $|\mock_i| \ne |\actual|$. 
\end{enumerate}
 Standard metrics for comparing ranked lists begin with correlation coefficients, like Kendall's $\tau$ or Spearman's rank correlation. However, these fail to address the three aforementioned problems. First, they are not weighed: differences at the end of the list are considered equal to discrepancies at the beginning of the list.\footnote{There are weighted versions of Kendall's $\tau$ from \cite{weightedkendallstau} that can address the weighting issue, but the two lists must still have the same length and elements.} Second, they must allows for lists of different lengths.\footnote{\cite{LinAggregationMethods} shows how Kendall's and Spearman's measures can be modified for incomplete lists.}

 Another possible metric is mean absolute error of the predicted and actual draft positions \citep{Insider, Samford}: 
 $$error(\mock_i,\actual) = \frac{1}{|\allplayers^*|} \sum_{p \in \allplayers^*} |\mock_i(p) - \actual(p)|$$
 where $\allplayers^* \subseteq \allplayers$. This is not front-weighted, but as mentioned before, this can be fixed with logarithms or similar transformations: 
 $$error(\mock_i,\actual) = \frac{1}{|\allplayers^*|} \sum_{p \in \allplayers^*} |log[\mock_i(p)] - log[\actual(p)]|.$$ 
 However, this still struggles with metric requirements 2 and 3. Undrafted players have no clear numerical value of pick. \cite{Harvard} imputes 61, and we have used 65 and 66 before ourselves. All are fairly arbitrary, and numbers close to $|\actual|$ over-penalize longer mock drafts/rankings where $|\mock_i| >> |\actual|$. Regarding unequal length, the selection of $\allplayers^*$ is nontrivial and not obvious. It should probably be all players, but in practice only the players actually drafted or the players on the mock draft are used. Instead, we propose a metric that more naturally handles incomplete lists of different lengths. 


\subsection{Rank-Biased Overlap/Distance}

Rank-biased overlap (RBO) is introduced by \cite{RBO} and addresses the aforementioned three issues. 
 The core idea of RBO is to measure the overlap between the set of selected players at each pick in the draft. It is ``rank-biased'' as earlier picks will be counted in every step afterward. 

Consider two complete lists: both an actual draft $\actual$ and a mock draft $\mock$ that orders an infinite number of players.
Let $X_d \in \{0,1,...,d\}$ be the number of players selected in both the mock and actual draft through pick $d$, such that $X_d = |\mock^{-1}(\{1,...,d\}) \cap \actual^{-1}(\{1,...,d\})|$, the size of the intersection between the first $d$ players in each list. We can then think of $\frac{X_d}{d}$ as the fraction or percent of agreement between the two lists through pick $d$.  RBO is a weighted average of $\frac{X_d}{d}$ over $d=1,...,\infty$ 
\begin{equation}
\label{eq:RBO}
RBO(\mock,\actual,q) = \frac{1-q}{q} \sum_{d=1}^{\infty} \frac{X_d}{d} q^{d}. 
\end{equation}
This  yields a measure of similarity between the lists that is weighed toward the top of the draft. For example, the first overall pick impacts every $X_d$, while the 31st pick in the draft does not impact $X_d$ for $d=1,...,30$.

The tuning parameter $q$ serves two functions.\footnote{\cite{RBO} use $p$ as the tuning parameter, but we call it $q$ to avoid confusion with $p$ as the index for players.}
First, for $0<q<1$, the weighted sum is constrained to the [0,1] interval and is thus comparable to correlation. Second, it allows us to adjust the degree of front weighting. For our purposes,  we chose $q=0.98$. This implies about 89\% of the metric's weight is on the 60 picks in the draft (thus 11\% to those who are undrafted), which includes 51\% of the weight on the lottery/the first 14 picks (also known as the ``lottery'').\footnote{Calculating the amount of weight for each draft position is not as simple as working with the geometric series. See Section 4.3 of \cite{RBO} for further details.}

However, Equation \ref{eq:RBO} only addresses the weighting criterion for our metric, not the incompleteness or nonequal-length criteria. 
To achieve this, \cite{RBO} consider taking finite lists and extrapolating to what they would do if infinite. Assume that the two lists may be of different length, where the longer list has $\ell$ elements and the shorter has $s$ elements. \cite{RBO} define extrapolated RBO, or $RBO_{EXT}$, which we write a bit differently as  
\begin{equation}
    \widehat{RBO}(\mock,\actual,p) =  
    \left\{\frac{1-p}{p}   \sum_{d=1}^{s} p^d \frac{X_d}{d}   \right\}
    + 
    \left\{\frac{1-p}{p}  \sum_{d=s+1}^{l} p^d \left[ \frac{X_d}{d} + \frac{X_s}{s} \left(1-\frac{s}{d}\right) \right]  \right\}
    +
    \left\{p^\ell\left(\frac{X_\ell}{\ell} + \frac{X_s}{s}\left[1-\frac{s}{\ell}\right]\right)\right\}
\label{eq:RBOEXT}
\end{equation}
The first summand in Equation \ref{eq:RBOEXT} is the same as Equation \ref{eq:RBO} but with only a finite sum: we only fully observe the overlap $X_d/d$ for the first $s$ elements as that's the length of the shorter list. Hence the sum goes from $d=1,...,s$.  

The second summand allows us to consider lists of different lengths (i.e. mock drafts that are not the same length as the actual draft). It also contains $\frac{X_d}{d}$, but here these are the observed agreement of the parts the longer list with the observed parts of the shorter list: only half of the information we want in $d=(s+1),...,\ell$. The other term in square braces assumes that the observed agreement of the short list is a good estimate of the agreement of the unobserved pieces.  

The third summand allows us to measure incomplete (i.e. finite) lists. It is the same as the second, except it now uses $\frac{X_\ell}{\ell}$ as the estimate of agreement for the unobserved elements of the longer list.  
Furthermore, $q$ weights the different picks in the draft, which weights become very small because $q^d \rightarrow 0$ as $d$ becomes large.

Rank-biased distance is the corresponding error metric, and similar to \cite{RBO}, we define it as
\begin{equation}
    RBD = 1 - \widehat{RBO}
\end{equation}
using this ``extrapolated'' version of RBO. 

\section{Rank aggregation for combining mock drafts}

\subsection{Standard Approach: Borda Count}

With an appropriate error metric, we can compare the mock draft accuracy of various authors and combination methods. 
The most common rank aggregation methods in sports are the Borda count or variations thereof. It is used for the aggregate/consensus mock drafts at both NBA.com's  and HoopsHype.com (Sports Illustrated and others reference this one). It is also used for the AP top 25 rankings of NCAA sports teams using the rankings provided by the AP members, and a similar system is used to combine votes for the annual NBA awards, including most-valuable player (MVP).
The basic idea is that each player receives points according to his forecasted draft slot. Then the mock drafts are aggregated by sorting players by how many total points they have obtained across the different mock drafts. 

For player $p$, assign a score by adding points from each author: 
$$BORDA_p =   \sum_{i=1}^{n_{author}} [|\actual|-\hat\mock_i(p)+1].$$
For the NBA draft $|\actual|=60$ usually, and players receive 60 points for each author that forecasts them to be drafted first overall, 59 points for second, and so forth, up to players forecasted to be undrafted then receiving zero points. To do such, we define
$$
\hat{\mock}(p)=
\begin{cases}
{\mock}(p),& {\mock}(p) \ne \undrafted\\
|\actual|+1 ,& {\mock}(p) = \undrafted
\end{cases}
$$
The aggregate/consensus mock draft is then created by ordering $BORDA_p$,  $\forall p$.\footnote{Simply averaging a player's forecasted draft slots and sorting on these average slots from smallest to largest is a special case of Borda counting. However, while just taking averages is simpler, players must appear in each of the mock drafts, otherwise that mock draft's contribution is not defined. Adding points and the sorting largest to smallest (Borda counting) is robust to this as being not listed by an author simply results in 0 points added.}


\subsubsection{Example: using the Borda count to aggregate rankings}
Assume we have the five ``greatest of all time'' (GOAT) mock drafts (i.e. rankings of top five best players ever) in Table \ref{tab:examplemocks}. Note that Mocks A through D are loosely taken from popular websites. Mock E on the other hand is ordered by the most championships after the NBA-ABA merger, with ties split by total points scored; this is an extreme version of saying that winning championships is most important. 

\begin{table}
    \centering
\begin{tabular}{c|lllll}
Pick & Mock A	&	Mock B	&	Mock C	&	Mock D	&	Mock E	\\
\hline
1 & M. Jordan	&	M. Jordan	&	M. Jordan	&	M. Jordan	&	R. Horry	\\
2& L. James	&	L. James	&	L. James	&	L. James	&	K. Abdul-Jabbar	\\
3&K. Abdul-Jabbar	&	K. Abdul-Jabbar	&	B. Russell	&	B. Russell	&	M. Jordan	\\
4&B. Russell	&	B. Russell	&	K. Abdul-Jabbar	&	K. Bryant	&	S. Pippen	\\
5&K. Bryant	&	E. Johnson	&	E. Johnson	&	K. Abdul-Jabbar	&	K. Bryant	\\
\end{tabular}
    \caption{Five example top 5 ``greatest of all time'' rankings.}
    \label{tab:examplemocks}
\end{table}

 Michael Jordan is ranked first overall four times, worth five points each, and third once, worth three points, giving Michael Jordan a score of 23 (not intentional). Doing the same calculations for each player, we get the aggregate ranking found in Table \ref{tab:exampleagg}.

\begin{table}
\centering
\begin{tabular}{llr|rl}																				
	  	\multicolumn{3}{c|}{Borda Count}	  &	\multicolumn{2}{c}{Ranked-Choice Aggregation}			 	\\
	\hline																			
	Rank	&	Name		&	Points	&	Rank	&	Name			\\
   \hline
	1	&	Michael	Jordan	&	23	&	1	&	Michael	Jordan		\\
	2	&	LeBron	James	&	16	&	2	&	LeBron	James		\\
	3	&	Kareem	Abdul-Jabbar	&	13	&	3	&	Kareem	Abdul-Jabbar		\\
	4	&	Bill	Russell	&	10	&	4	&	Bill	Russell		\\
	5	&	Robert	Horry	&	5	&	5	&	Kobe	Bryant		\\
	6	&	Kobe	Bryant	&	4	&	6	&	Earvin	``Magic"	Johnson	\\
	7 (tie)	&	Earvin ``Magic"	Johnson	&	2	&	7	&	Robert	Horry		\\
	7 (tie)	&	Scottie	Pippen	&	2	&	8	& Scottie Pippen		\\
	\hline																			
\end{tabular}			
\caption{The results of combining the example rankings in Table \ref{tab:examplemocks} using both Borda count and RCA methods. 
}
\label{tab:exampleagg}
\end{table}			

\pagebreak
One downside of Borda count is that an outlier mock draft, like Mock E here, can have an out-sized impact in the result. Issues like this are known to be a problem in the literature on voting methods, e.g. \cite{bordaissue}.
Few fans would probably rank Robert Horry in their top five players ever, but in this example he does in fact land in the top five. This can happen with the actual mock drafts as well. Including more mock drafts and obtaining longer lists can both help with this issue, but moving away from a scoring system like this can help this issue even for short mock drafts. Of course, one could simply remove outlier mock drafts, but perhaps there is information contributed by that author that you would like to keep. 

\subsection{Ranked-Choice Aggregation}

There are many methods for aggregating rankings, such as those covered in \cite{LinAggregationMethods}, but for the purposes of the sports world, the method should be conceptually comprehensible by analysts and fans alike. Ranked-choice voting is a voting system where voters' ballots rank candidates from most-preferred to least-preferred. The benefit is that candidates can express more of their opinion than one single option. Sometimes called instant-runoff voting, it is growing in usage in elections around the world and thus provides an opportunity for us employ it as audiences grow in familiarity. 
Roughly speaking, ranked-choice voting has the following steps:
\begin{enumerate}
    \item Count each voters' most preferred candidate as the votes. 
    \item If one of the candidates has a majority, that candidate wins. Else, go to step 3
    \item Drop the candidate with the fewest votes from all voters' ranked lists. Return to step 1. 
\end{enumerate}

If we assume that mock drafts are ranked lists, then with a few tweaks we can use the idea of ranked-choice voting to combine mock drafts. The main needed tweak is that we go through the ranked-choice voting process for each pick in the draft, and after a player is picked, they are obviously removed from ballots for all future picks. The idea is conceptually mentioned in \cite{VOI}, however here we will detail these steps in Algorithm \ref{algo:RCAdraft}, which we call ``ranked-choice aggregation" or RCA. Note that lines 4-9 and 12-16 are the original bones of ranked-choice voting.

\begin{algorithm}
        \caption{Ranked-choice aggregation for combining mock drafts. Note that $P$ and $D$ are different than $\allplayers$ and $\actual$. Here, $P$ can only be all the players in a mock draft, such that $P = \cup_i \mock_i$. Players that are draft eligible but do not appear on any mock draft have no information and are thus left off the final list. $D$ is the number of picks the user would like the algorithm to produce, but $D \le |P|$. $D=\actual$ is a reasonable/usual choice.  $*$ In the event that an author published a mock draft with tiers, then players would be tied and each player counted in line 5. $**$ If there is a tie for fewest votes, the player with the highest average mock draft position is selected to be dropped in line 13.} 
        \label{algo:RCAdraft}
        \begin{algorithmic}[1]        
    \State $P^{\bullet} \gets P$ \Comment{set the list of available players $P^{\bullet}$ equal to the full list of players $P$}
    \While{$d \leq D$}
        \State $P^{\bullet\bullet} \gets P^\bullet$ \Comment{set the list of all players eligible for pick $d$, $P^{\bullet\bullet}$, equal to  $P^{\bullet}$}
        \While{$\tilde{\mock}_{RCA}^{-1}(d)$ undefined}  
            \State $ballot_i \gets \argmin_{p \in P^{\bullet\bullet}}\mock_i(p)$, $\forall i$ \Comment{each author votes for their highest ranked eligible player$.^*$ }
            \State $votes_p \gets |\{i:ballot_i = p\}|$, $\forall p\in P^{\bullet\bullet}$, \Comment{count the number of votes for each player}
            \State $\tilde{p} \gets \argmax_p votes_p$ \Comment{find the player with the most votes}
            \If{$votes_{\tilde{p}} > \frac{1}{2}n_{authors}$} \Comment{if it is a majority}
                \State $\tilde{\mock}_{RCA}(\tilde{p}) \gets d$ \Comment{Set the RCA mock to draft player $\tilde{p}$ with pick $d$}
                \State drop $\tilde{p}$ from $P^\bullet$  \Comment{Player $\tilde{p}$ is no longer available}
                \State $d \gets d+1$
            \Else 
                \State $p^* \gets \argmin_{p:votes_p>0} votes_p$  \Comment{find the player with the fewest, nonzero votes$.^{**}$}
                \State drop $p^*$ from $P^{\bullet\bullet}$  \Comment{Player $p^*$ is no longer eligible to be drafted with pick $d$}
            \EndIf
        \EndWhile
\EndWhile
\end{algorithmic}

\end{algorithm}


The step-by-step workings of ranked-choice aggregation on the GOAT example is in the Appendix \ref{app:RCA}, but for parsimony, it suffices us to produce the ranked-choice aggregation combined ranking in the right panel of Table \ref{tab:exampleagg}.  We submit that these rankings better reflect the overall opinions represented in the original mock drafts than do the traditional version via Borda counting.  Though many fans would rank Pippen over Horry (and would put neither in the top 8 all time), that is not seen in the example mocks, but if it is true it would be seen in the aggregate if we had longer ranked lists from each author or lists from more authors.

\section{Data}

\begin{table}
\centering
\begin{tabular}{r|rrr|rrrr|r}																				
	\hline																			
				&	 	\multicolumn{3}{c|}{Authorship Counts}	  &	\multicolumn{4}{c|}{Length of Mock Draft}			&	\\
			Year	&	Authors	&	Authors*Type	&	$\le 30$ Days 	&	$\le 14$	& 15-30	&	 31-60	&  $>60$	& 	Mock Drafts	\\
   \hline
2009	&	5	&	5	&	5	&	1	&	3	&	1	&	0	&	5	\\
2010	&	17	&	17	&	11	&	0	&	10	&	7	&	0	&	17	\\
2011	&	12	&	12	&	9	&	1	&	7	&	4	&	0	&	12	\\
2012	&	10	&	10	&	8	&	3	&	4	&	3	&	0	&	10	\\
2013	&	15	&	16	&	15	&	3	&	9	&	4	&	1	&	17	\\
2014	&	16	&	16	&	8	&	2	&	9	&	3	&	2	&	16	\\
2015	&	24	&	24	&	15	&	3	&	18	&	7	&	0	&	28	\\
2016	&	14	&	14	&	8	&	1	&	11	&	3	&	1	&	16	\\
2017	&	44	&	48	&	37	&	3	&	26	&	20	&	14	&	63	\\
2018	&	108	&	120	&	67	&	7	&	153	&	56	&	46	&	262	\\
2019	&	173	&	194	&	70	&	47	&	267	&	102	&	66	&	482	\\
2020	&	77	&	83	&	22	&	5	&	93	&	45	&	28	&	171	\\
2021	&	73	&	76	&	37	&	1	&	104	&	57	&	4	&	166	\\
	\hline																			
\end{tabular}			
\caption{Counts of the number of mock drafts by year and number of authors by year. Column 2 shows the number of different authors by year. Column 3 shows the number of different sources, for example Author A may publish two types of mock draft, an opinion based on A's expertise and a mock based on a statistical model. Column 4 count the number of Authors*Type that put out a mock in the 30 days prior to the draft; these are the point plotted in Figure \ref{fig:years}. Columns 5-8 tabulate the number of mocks by their length: lottery-only ($\le 14$ picks in mock), the first round (15-30), two rounds (31-60), or a ``big board'' of eligible players ($> 60$). Length is determined by the maximum rank value (AWK). The final column is the total number of mock drafts in the database in a given year.}
\label{tab:data}
\end{table}																				

Our dataset is the largest known database of NBA mock drafts, and contains of mock drafts from over 100 different authors at different points in time from the 2009 draft to the 2021 draft. Table \ref{tab:data} tabulates the number of authors and mock drafts collected by draft year, and the final column shows that, generally, more were collected as time went on. Data collection began after the 2017 NBA draft, such that  nearly every mock draft on the internet was collected for the 2018 and 2019 NBA drafts.   The years following 2019 still have many mock drafts, though the set of authors has become more refined to  trustworthy 
sources. Mock drafts prior to 2017  were collected using online archives.

Table  \ref{tab:data} also shows that typically we have more mock drafts than authors in a given year. Naturally, serious authors publish multiple mock drafts over the course of the year as more information becomes available about player ability and team preferences. If we need a single mock draft from an author for a particular season, we use their last mock draft of the season.   

Finally, the middle columns of Table \ref{tab:data} shows how long each of these mock drafts are. Often, mock drafts only forecast the first round of the draft (up to 30 picks, one per team). This is a bit surprising given the fact that there are two rounds. However, the second round is much harder to forecast accurately for many reasons, including that player abilities are more similar later in the draft, and because errors in the draft tend to propagate more errors later in the draft (e.g. a player being picked earlier than expected may cause all following players to be picked a slot later than forecasted, and so forth).

\section{Results}

\begin{figure}
    \centering
        \includegraphics[width=5.88in]{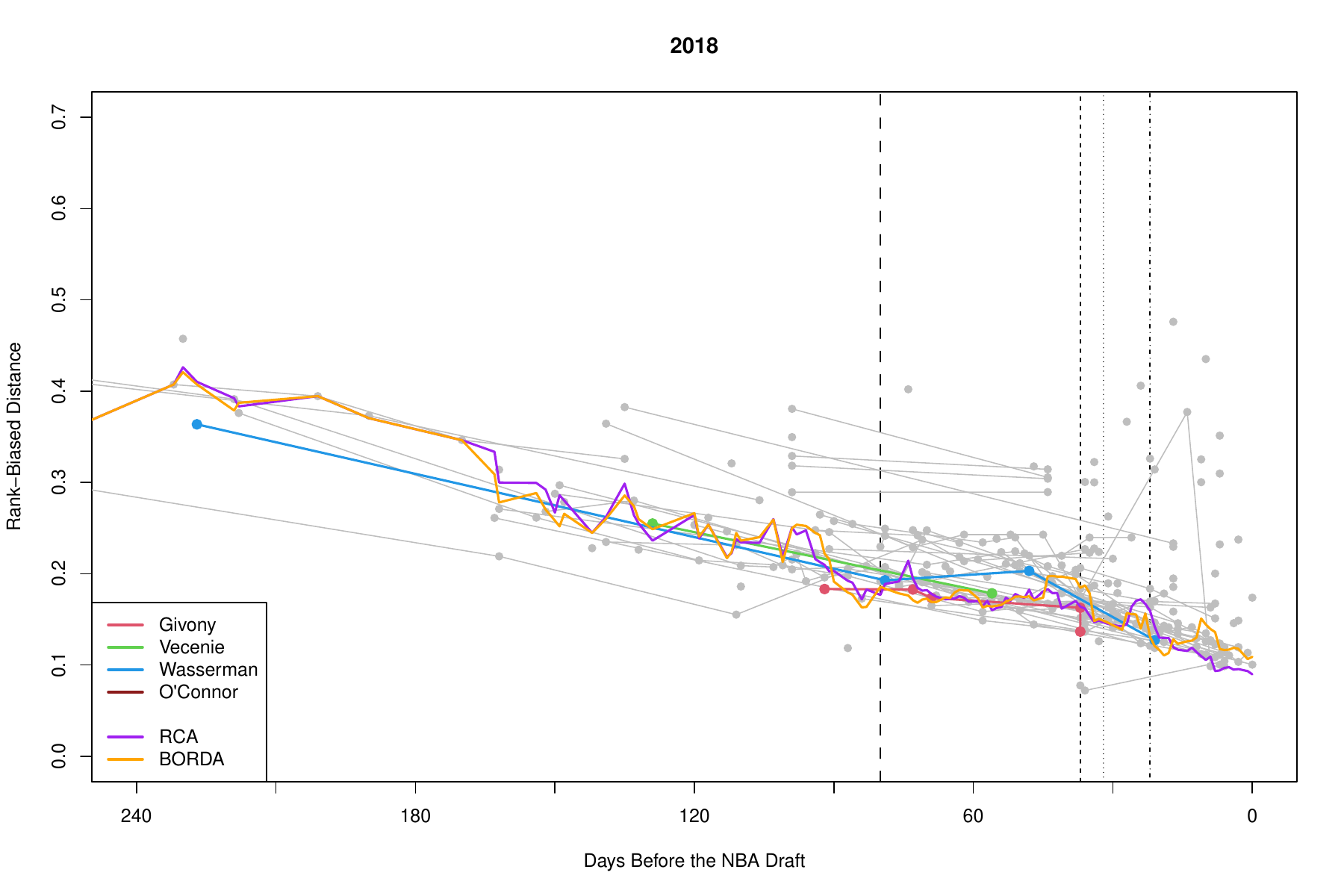}\\
        \includegraphics[width=5.88in]{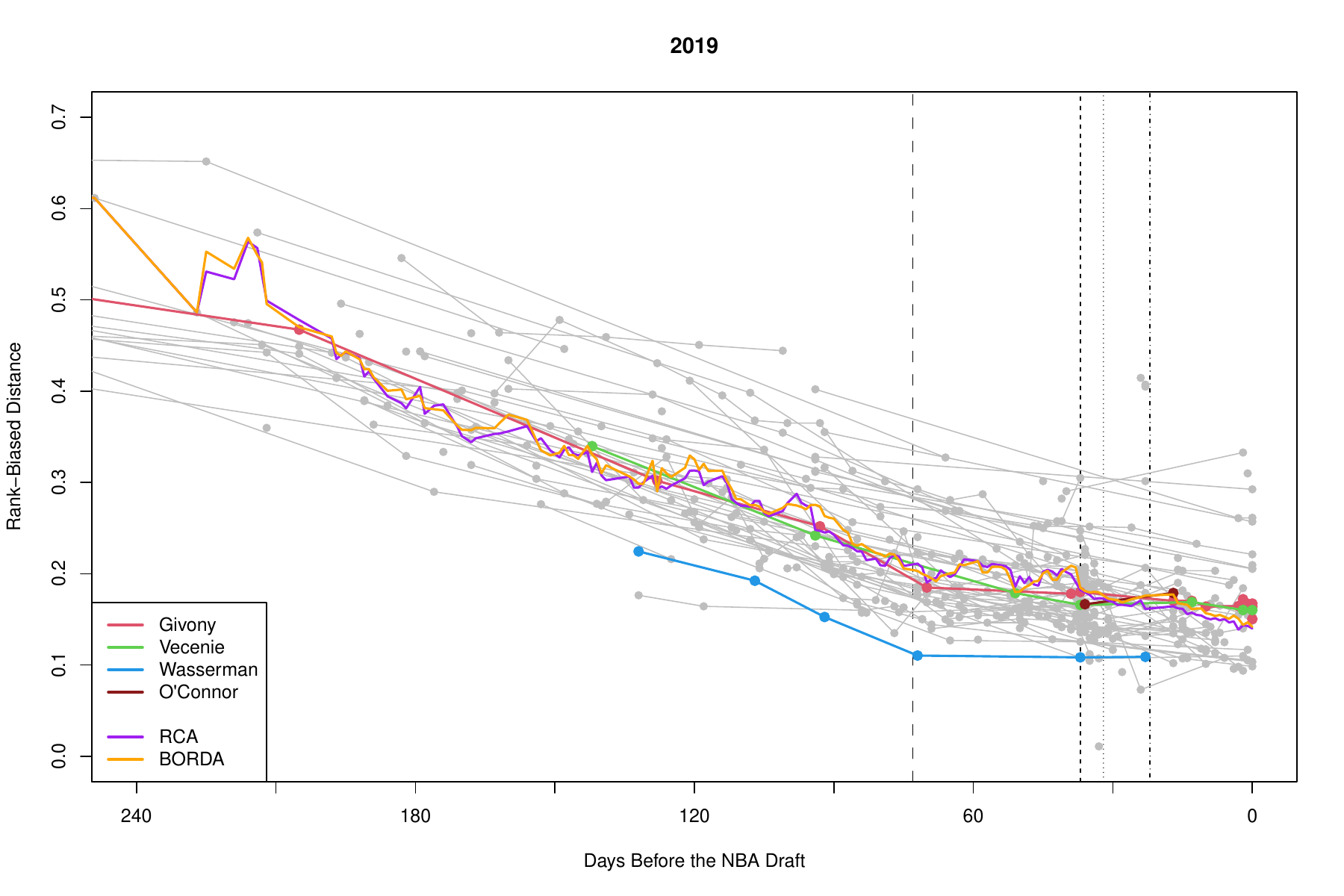}
    \caption{Mock draft error over time within each season.  Select authors are denoted by color: Jonathan Givony in red, Sam Vecenie in green,  Johnathan Wasserman in blue, and Kevin O'Connor in brown. The two aggregation methods' performances are also shown in color:  RCA in purple and Borda-counting in orange.  Important pre-draft dates are shown with vertical lines: large dashes = NCAA Championship Game, small dashes = NBA Draft Lottery, dotted = last day of NBA Draft Combine, dot-dashed = NCAA withdrawal deadline.}
    \label{fig:days}
\end{figure}

We look at mock draft error over time, specifically 1) within season for the days leading up to the draft, and 2) across seasons. Figure \ref{fig:days} shows mock draft errors in the days leading up to the NBA Draft, for four chosen seasons. Four well-known authors (e.g. NBA.com uses them in its consensus mock draft) are highlighted: ESPN's Jonathan Givony in red, The Athletic's Sam Vecenie in green, Bleacher Report's Jonathan Wasserman in blue, and The Ringer's Kevin O'Connor in brown. We see that these authors are consistently among the best of the authors in terms of overall forecast error. It is interesting to note that in 2019, we see an interesting kink in the trends corresponding to the end of the NCAA tournament. Prior to this, there was rapid improvement of all mock drafts in general, but afterward there was little change in terms of overall accuracy. 

\begin{figure}
    \centering
    \includegraphics[width=6.5in]{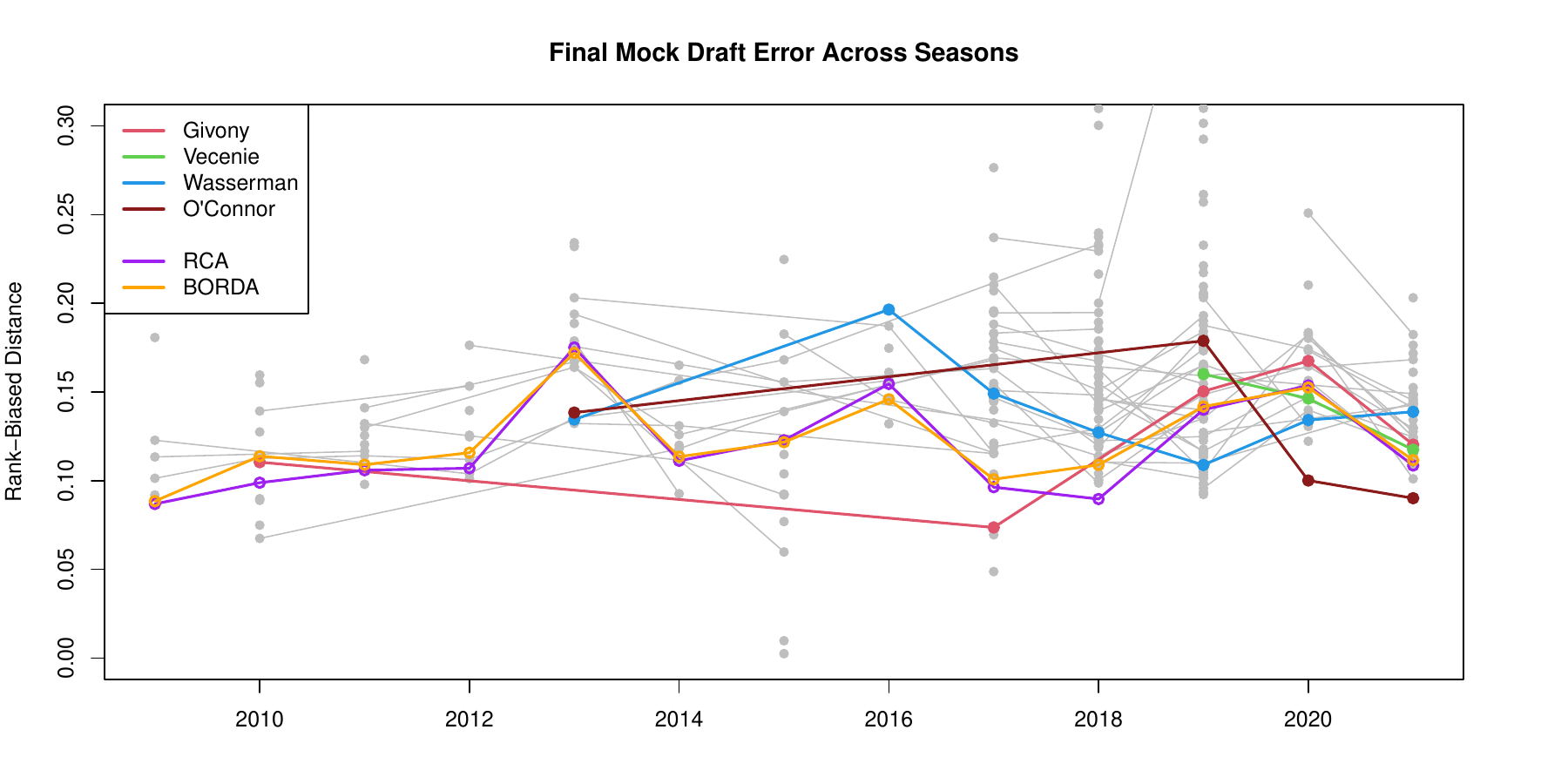}
    \caption{Accuracy of the each author's last mock draft before the actual draft, for select authors, season to season. If their final mock was within 30 days of the draft. Select authors are denoted by color: Jonathan Givony in red, Sam Vecenie in green,  Johnathan Wasserman in blue, and Kevin O'Connor in brown. The two aggregation methods' performances are also shown in color:  RCA in purple and Borda-counting in orange. }
    \label{fig:years}
\end{figure}

Also highlighted in the plot are the two aggregation methods we have described: BORDA in orange and RCA in purple. 
When doing in-season RCA or BORDA over time, the aggregated mock draft $d$ days before the actual drafts only looks at mock drafts that were posted in the past 10 days, and only use each author's most recent mock draft. 
Either combination method is more consistently in the group of lowest-error mocks than any single author. 
Furthermore, the two combination methods alternate as to which is more accurate at any given point in a season. However, we see that RCA performs better than BORDA (purple line is beneath the orange line) at the end of both the 2018 and 2019 seasons, when draft forecast accuracy is most important. Indeed, Figure \ref{fig:years} shows that this is the case in most every season. 



Figure \ref{fig:years} shows mock draft error across seasons. 
The points included are each author's final mock draft within 30 days before the draft. 
We see that mock drafts right before the actual draft are consistently fairly accurate, but do have error in the 0.1 to  0.2 range. 
Again, we see that the three highlighted authors tend to be among the best authors (hence earning their standing as some of the best). 
We also note that, again, RCA regularly provides a forecast similar to or better than BORDA's, with exceptions only in 2013 and 2016.   

These trends are examined numerically in Table \ref{tab:percentiles}, where we report the percentile rankings of regular authors' final mock drafts over the seasons.   Regular authors are defined as those with three or more seasons' mock drafts in the dataset. Final mock drafts are defined as the last mock draft produced by a particular author in the 30 days preceding the actual draft, including the day of the actual draft. 
We learn that, on average, RCA will provide a final mock draft in the 74th percentile, while BORDA will be in the 68th percentile. This places RCA in third place among these 26 regular authors, and BORDA at sixth. Of note are the two authors ahead of RCA, Kyle Irving and Kevin O'Connor, whose percentiles are bolstered by having only a few seasons in the dataset but with excellent forecast performance.  
We conclude that the ideal way to produce reasonably accurate forecasts of the NBA draft is to collect mock drafts from the best authors and combine them with RCA. 


\begin{sidewaystable}[ht]
\centering
\begin{tabular}{lrrrrrrrrrrrrrc}
  \hline
 Author & 2009 & 2010 & 2011 & 2012 & 2013 & 2014 & 2015 & 2016 & 2017 & 2018 & 2019 & 2020 & 2021 & Avg Perc \\ 
  \hline
Kyle Irving  &  &  &  &  &  &  &  &  &  &  & 0.97 & 0.78 & 0.74 & 0.83 \\ 
Kevin O'Connor  &  &  &  &  & 0.81 &  &  &  &  &  & 0.32 & 1.00 & 1.00 & 0.78 \\ 
\bf{RCA} & 1.00 & 0.67 & 0.90 & 0.78 & 0.44 & 0.89 & 0.38 & 0.67 & 0.92 & 1.00 & 0.62 & 0.52 & 0.95 & \bf{0.75} \\ 
Chad Ford (R) &  &  &  &  & 1.00 & 0.22 &  &  & 0.84 &  &  &  &  & 0.69 \\ 
Zach Buckley  &  &  &  &  &  &  &  &  &  & 0.90 & 0.82 &  & 0.34 & 0.69 \\ 
\bf{BORDA} & 0.83 & 0.42 & 0.80 & 0.56 & 0.50 & 0.67 & 0.44 & 0.78 & 0.89 & 0.91 & 0.55 & 0.61 & 0.87 & \bf{0.68} \\ 
Sam Vecenie  &  &  &  &  &  &  &  &  &  &  & 0.45 & 0.74 & 0.76 & 0.65 \\ 
Chad Ford  & 0.17 &  &  & 0.89 & 0.88 &  &  & 0.56 &  &  &  &  &  & 0.62 \\ 
Jonathan Givony  &  & 0.58 &  &  &  &  &  &  & 0.95 &  & 0.49 & 0.30 & 0.71 & 0.61 \\ 
Jonathan Wasserman  &  &  &  &  & 0.94 &  &  & 0.00 & 0.55 & 0.65 & 0.83 & 0.87 & 0.45 & 0.61 \\ 
Ricky O'Donnell  &  &  &  &  &  &  &  &  & 0.53 & 0.46 & 0.76 & 0.70 &  & 0.61 \\ 
Joe Kotoch  &  &  & 0.30 &  & 0.62 &  & 0.88 &  &  &  &  &  &  & 0.60 \\ 
Brad Rowland  &  &  &  &  &  &  &  &  &  & 0.37 & 0.89 & 0.09 & 0.97 & 0.58 \\ 
Jonathan Wasserman (R) &  &  &  &  &  &  &  &  & 0.76 & 0.63 & 0.35 &  &  & 0.58 \\ 
Gary Parrish  &  &  &  &  &  &  & 0.12 & 0.89 & 0.71 & 0.82 & 0.63 & 0.13 & 0.63 & 0.56 \\ 
Jeremy Woo  &  &  &  &  &  &  &  &  &  & 0.84 & 0.51 & 0.35 & 0.55 & 0.56 \\ 
Jordan Schultz  &  &  & 0.20 &  &  & 0.78 & 0.69 &  &  &  &  &  &  & 0.56 \\ 
Kyle Boone  &  &  &  &  &  &  &  &  &  & 0.66 & 0.39 &  & 0.61 & 0.55 \\ 
My NBA Draft  &  & 1.00 &  &  &  & 0.56 & 0.31 &  & 0.37 &  &  &  & 0.18 & 0.48 \\ 
Adi Joseph  &  &  &  &  & 0.69 & 0.33 &  &  & 0.39 &  &  &  &  & 0.47 \\ 
Sean Deveney  &  &  &  &  & 0.38 &  & 0.25 &  & 0.45 & 0.78 &  &  &  & 0.47 \\ 
Paul Banks  & 0.50 & 0.50 & 0.70 & 0.67 &  & 0.11 & 0.19 &  &  & 0.18 &  &  &  & 0.41 \\ 
Scott Howard-Cooper  &  &  &  &  & 0.12 &  &  & 0.11 & 0.82 &  &  &  &  & 0.35 \\ 
Steve Kyler  &  &  &  &  &  &  &  &  & 0.34 & 0.43 & 0.28 &  &  & 0.35 \\ 
NBA Draft Room  &  &  &  &  &  &  &  &  & 0.08 & 0.53 & 0.24 &  &  & 0.28 \\ 
Tankathon  &  &  &  &  &  &  &  &  &  &  & 0.27 & 0.22 & 0.29 & 0.26 \\ 
   \hline
\end{tabular}
\caption{Regular authors' percentile rankings of their final mock drafts. Regular authors are defined as those with three or more seasons' mock drafts in the dataset. Final mock drafts are defined as the last mock draft produced by a particular author in the 30 days preceding the actual draft, including the day of the actual draft. Percentile of 1.0 indicates the best mock draft in the set, and 0 is the worst. The table is sorted by the final column, which is the average percentile of the row/across the seasons. Authors with ``(R)'' notation indicate players rankings produced by that author, as opposed to their mock drafts.}
\label{tab:percentiles}
\end{sidewaystable}


\section{Conclusion}

We assume the NBA draft is a ranked list and submit that this is reasonable to assume as NBA teams usually draft the most talented player regardless of positional fit. Under this assumption, commonly-used metrics of mock draft accuracy/error are flawed: namely, they are not weighted toward the top of the draft, require mock drafts to be of a certain length, and need mock drafts to contain the same players as the actual draft. Some proposed metrics fix some of the problems, but not all. Thus, we propose rank-biased distance (RBD) of \cite{RBO} as the appropriate metric. 

Using  the largest known database of mock drafts, we show that the well-known mock draft authors are more accurate in terms of RBD than the average mock draft on the internet. Furthermore, mock drafts tend to become more accurate as the actual draft approaches, as one would expect. Perhaps unexpectedly, mock draft accuracy in the week or so before the actual draft tends to be the same year to year. 

We also propose that ranked-choice aggregation be used to combine mock drafts into a single consensus mock draft. While there are many aggregation methods available in literature on set theory, voting theory, and others, we propose that basing aggregation on the ideas of ranked-choice voting is more approachable for the analysts that create consensus mock drafts and the audience that consumes them. Perhaps changes can come and Borda count will either be used less or be used more carefully, as we see that ranked-choice aggregation is as accurate or more accurate than the industry standard Borda count.




\bibliographystyle{ba}
\bibliography{bib}


\newpage
\appendix
\section{Appendix: Detailed RCA demonstration}    
\label{app:RCA}

To demonstrate how ranked-choice aggregation works, we now show how to aggregate the GOAT mock drafts of Table \ref{tab:examplemocks} using Algorithm \ref{algo:RCAdraft}. The first set of ballots are simply the top ranked players from each of the mocks (we can just read across row 1), such that there are four votes for Michael Jordan and one vote for Robert Horry, so that Jordan is clearly the first pick of aggregate mock, and is removed from the list of available players. We can then visualize the rankings as 
\\
\vspace{12pt}
\\
\begin{tabular}{c|lllll}
Order & Mock A	&	Mock B	&	Mock C	&	Mock D	&	Mock E	\\
\hline
1	&	L. James	&	L. James	&	L. James	&	L. James	&	R. Horry	\\
2	&	K. Abdul-Jabbar	&	K. Abdul-Jabbar	&	B. Russell	&	B. Russell	&	K. Abdul-Jabbar	\\
3	&	B. Russell	&	B. Russell	&	K. Abdul-Jabbar	&	K. Bryant	&	S. Pippen	\\
4	&	K. Bryant	&	E. Johnson	&	E. Johnson	&	K. Abdul-Jabbar	&	K. Bryant	\\
\end{tabular}
\\
\vspace{12pt}
\\
and the ballots for the second pick are counted with four votes for LeBron James and one vote for Robert Horry. Thus, LeBron James is clearly the second pick of the aggregate mock and is removed for the list of available players:
\\
\vspace{12pt}
\\
\begin{tabular}{c|lllll}
Order & Mock A	&	Mock B	&	Mock C	&	Mock D	&	Mock E	\\
\hline
1	&	K. Abdul-Jabbar	&	K. Abdul-Jabbar	&	B. Russell	&	B. Russell	&	R. Horry	\\
2	&	B. Russell	&	B. Russell	&	K. Abdul-Jabbar	&	K. Bryant	&	K. Abdul-Jabbar	\\
3	&	K. Bryant	&	E. Johnson	&	E. Johnson	&	K. Abdul-Jabbar	&	S. Pippen	\\
4	&		&		&		&		&	K. Bryant	\\
\end{tabular}
\\
\vspace{12pt}
\\
For the third aggregate RCA pick, the ballots yield two votes for Kareem Abdul-Jabbar, two votes for Bill Russell, and one vote for Robert Horry. As no player has a majority, we drop the player with the fewest positive number of votes from the list of eligible players, which here is Robert Horry. 
\\
\vspace{12pt}
\\
\begin{tabular}{c|lllll}
Order & Mock A	&	Mock B	&	Mock C	&	Mock D	&	Mock E	\\
\hline
1	&	K. Abdul-Jabbar	&	K. Abdul-Jabbar	&	B. Russell	&	B. Russell	&	\sout{R. Horry} 	\\
2	&	B. Russell	&	B. Russell	&	K. Abdul-Jabbar	&	K. Bryant	&	K. Abdul-Jabbar	\\
3	&	K. Bryant	&	E. Johnson	&	E. Johnson	&	K. Abdul-Jabbar	&	S. Pippen	\\
4	&		&		&		&		&	K. Bryant	\\
\end{tabular}
\\
\vspace{12pt}
\\
Now recounting the ballots (i.e. each author's top eligible player), we have three votes for Kareem Abdul Jabbar, which is a majority, making him the RCA aggregate mock's third pick and removing him from the available list. This leads to the ballots for the fourth pick
\\
\vspace{12pt}
\\
\begin{tabular}{c|lllll}
Order & Mock A	&	Mock B	&	Mock C	&	Mock D	&	Mock E	\\
\hline
1	&	B. Russell	&	B. Russell	&	B. Russell	&	B. Russell	&	R. Horry	\\
2	&	K. Bryant	&	E. Johnson	&	E. Johnson	&	K. Bryant	&	S. Pippen	\\
3	&		&		&		&		&	K. Bryant	\\
\end{tabular}
\\
\vspace{12pt}
\\
which is clearly won by Bill Russell. For the fifth pick
\\
\vspace{12pt}
\\
\begin{tabular}{c|lllll}
Order & Mock A	&	Mock B	&	Mock C	&	Mock D	&	Mock E	\\
\hline
1	&	K. Bryant	&	E. Johnson	&	E. Johnson	&	K. Bryant	&	R. Horry	\\
2	&		&		&		&		&	S. Pippen	\\
3	&		&		&		&		&	K. Bryant	\\
\end{tabular}
\\
\vspace{12pt}
\\
there is not an immediate majority, with Kobe Bryant and Magic Johnson each receiving two votes and Robert Horry receiving one. We deem Robert Horry ineligible for this pick as he has the fewest picks and recount the ballots. 
\\
\vspace{12pt}
\\
\begin{tabular}{c|lllll}
Order & Mock A	&	Mock B	&	Mock C	&	Mock D	&	Mock E	\\
\hline
1	&	K. Bryant	&	E. Johnson	&	E. Johnson	&	K. Bryant	&	\sout{R. Horry}	\\
2	&		&		&		&		&	S. Pippen	\\
3	&		&		&		&		&	K. Bryant	\\
\end{tabular}
\\
\vspace{12pt}
\\
There is still not an immediate majority, with Kobe Bryant and Magic Johnson each receiving two votes and Scottie Pippen receiving one. Now Scottie Pippen is dropped from the eligible list 
\\
\vspace{12pt}
\\
\begin{tabular}{c|lllll}
Order & Mock A	&	Mock B	&	Mock C	&	Mock D	&	Mock E	\\
\hline
1	&	K. Bryant	&	E. Johnson	&	E. Johnson	&	K. Bryant	&	\sout{R. Horry}	\\
2	&		&		&		&		&	\sout{S. Pippen}	\\
3	&		&		&		&		&	K. Bryant	\\
\end{tabular}
\\
\vspace{12pt}
\\
and we see that Kobe has three votes and Magic has two, earning Kobe the nod for the fifth pick of our aggregate mock draft. At this point, two of the mocks have used all of their listed picks, such that the next round of voting sees Magic Johnson with two votes and Robert Horry with one, giving Magic the sixth pick. That leaves one mock remaining, such that we simply follow its order. Hence, our RCA aggregate mock is as follows. 
 \begin{enumerate}
     \item Michael Jordan 
     \item LeBron James 
     \item  Kareem Abdul-Jabbar 
     \item Bill Russell 
     \item Kobe Bryant 
     \item Earvin ``Magic" Johnson 
     \item Robert Horry 
     \item Scottie Pippen 
 \end{enumerate}

\end{document}